\documentclass[submission,copyright,creativecommons]{eptcs}
\providecommand{\event}{SOS 2007} 

\usepackage{iftex}

\ifpdf
  \usepackage{underscore}         
  \usepackage[T1]{fontenc}        
\else
  \usepackage{breakurl}           
\fi

\usepackage{graphicx}
\usepackage{amsfonts}
\usepackage{caption}
\usepackage{subcaption}
\usepackage{footnote}
\usepackage{enumerate}
\usepackage{todonotes}
\usepackage{xcolor}
\usepackage{footnote}
\usepackage{xparse}
\usepackage{cite}
\usepackage{booktabs}
\usepackage{url}
\usepackage{algorithm}
\usepackage{algpseudocodex}
\hypersetup{breaklinks=true,
    colorlinks=true,       
    linkcolor=red,          
    citecolor=purple,        
    filecolor=purple,         
    urlcolor=magenta        
}
\usepackage{amssymb}
\usepackage{amsmath}
\usepackage{keyval}
\usepackage{xspace}
\usepackage{paralist}
\usepackage{listings}
\usepackage{multirow}
\usepackage{stmaryrd}
\usepackage{adjustbox} 
\usepackage{makecell}
\usepackage{sidecap}
\usepackage{booktabs} 
\usepackage{threeparttable, tablefootnote}
\usepackage{multirow}
\usepackage{xcolor}
\usepackage{comment}

\RequirePackage{tikz}
\usetikzlibrary{arrows,automata,shapes,calc,through,decorations.pathmorphing,decorations.fractals,chains,shapes.multipart}

\usepackage{arydshln} 
\usepackage[free-standing-units]{siunitx}
\usepackage{circuitikz}
\usepackage{tabularx} 

\newboolean{blueMode}
\setboolean{blueMode}{false}
\ifthenelse{\boolean{blueMode}}{
  
}
{
  
}

\title{Formal Verification of Long Short-Term Memory based Audio Classifiers: A Star based  Approach}
\author{Neelanjana Pal
\institute{Institute for Software Integrated Systems\\ Vanderbilt University\\ Nashville, USA}
\email{neelanjana.pal@vanderbilt.edu}
\and
Taylor T Johnson
\institute{Institute for Software Integrated Systems\\ Vanderbilt University\\ Nashville, USA}
\email{taylor.johnson@vanderbilt.edu}
}
\def\titlerunning{Formal Verification of Long Short-Term Memory based Audio Classifiers}
\def\authorrunning{N. Pal et al.}

\begin{document}
\maketitle
\newtheorem{definition}{Definition}[section]
\begin{abstract}
Formally verifying audio classification systems is essential to ensure accurate signal classification across real-world applications like surveillance, automotive voice commands, and multimedia content management, preventing potential errors with serious consequences. Drawing from recent research, this study advances the utilization of star-set-based formal verification, extended through reachability analysis, tailored explicitly for Long Short-Term Memory architectures and their Convolutional variations within the audio classification domain. By conceptualizing the classification process as a sequence of set operations, the star set-based reachability approach streamlines the exploration of potential operational states attainable by the system. The paper serves as an encompassing case study, validating and verifying sequence audio classification analytics within real-world contexts. It accentuates the necessity for robustness verification to ensure precise and dependable predictions, particularly in light of the impact of noise on the accuracy of output classifications.

\end{abstract}
\section{Introduction} 
\label{Sec: Introduction}
Deep Neural Networks (DNNs) have demonstrated remarkable capabilities in addressing intricate tasks like image classification, object detection, speech recognition, natural language processing, and document analysis, at times even surpassing human performance \cite{lawrence1997face, krizhevsky2012imagenet, lecun1998gradient}. This success has ignited a surge in exploring the viability of DNNs across diverse real-world domains, including biometric authentication, mobile facial recognition for security, and malware detection. However, given the sensitive nature of the data in these critical applications, incorporating safety, security, and robust verification into their design has become paramount.

However, studies have revealed that even slight modifications in input data can effectively mislead cutting-edge, well-trained networks, causing inaccuracies in their predictions \cite{moosavi2016deepfool, LBFGS, goodfellow2014explaining}. The arena of network verification has primarily concentrated on image inputs, particularly emphasizing the assurance of safety and robustness in various classification neural networks \cite{tran2021robustness, anderson2019optimization, botoeva2020efficient, katz2019marabou, mohapatra2020towards, tran2020verification}. Previous investigations have scrutinized a range of network architectures, encompassing feed-forward neural networks (FFNNs \cite{tran2019star}), convolutional neural networks (CNNs \cite{tran2020verification}), semantic segmentation networks (SSNs \cite{tran2021robustness}), and a few 
 using Recurrent Neural Networks (RNNs \cite{tran2023verification}) employing diverse set-based reachability tools such as Neural Network Verification (NNV \cite{tran2020nnv, lopez2023nnv}) and JuliaReach \cite{bogomolov2019juliareach}, among others.

Models utilizing NNs for audio classification have found application in diverse tasks, ranging from Music Genre Classification \cite{dong2018convolutional, choi2017convolutional, falola2022music} and Environmental Sound Classification \cite{guzhov2021esresnet, aytar2016soundnet, demir2020new} to Audio Generation \cite{oord2016wavenet, roberts2018hierarchical}. Therefore, formal verification of audio classification systems holds paramount importance in ensuring their reliability and safety, particularly in safety-critical applications such as autonomous vehicles \cite{walden2022improving, raja2021av}, medical diagnosis \cite{hemdan2023cr19, modegi1997application}, and industrial monitoring \cite{wang2003industrial}. 

This study introduces an extension, building upon the foundations laid by two recent studies \cite{tran2023verification, pal2023robustness} in the domain of formal verification. The objective is to leverage set-based reachability techniques to verify audio classification models based on the Long Short Term Memory (LSTM) and CNN-LSTM architectures. Drawing inspiration from \cite{tran2023verification}, which highlights the star-based verification of basic vanilla RNNs, and from \cite{pal2023robustness}, which demonstrates the formal verification of convolutional neural networks operating on time series data, work shown in this paper amalgamates both concepts. Specifically, it employs two LSTM models and one CNN-LSTM model for these classifications, following the ones depicted in \cite{mathworksClassifySound, mathworksSequenceClassification, mathworksSequenceClassification1d}.

\paragraph*{Contributions.}
\begin{enumerate}
    \item This paper presents a thorough case study on the formal verification of audio classification models using the LSTM and CNN-LSTM architectures with two different datasets. Our focus is to rigorously assess the robustness verification of these models within a formal verification framework, analyzing their behavior and performance against input noises. We develop our work as an extension of the NNV tool\footnote{The code for this paper is available at \url{https://github.com/verivital/nnv/tree/master/code/nnv/examples/Submission/FMAS2023}} to formally analyze and explore CNN-LSTM architecture verification for audio data using sound and deterministic reachability methods.
    \item Building on insights from existing research \cite{tran2023verification, pal2023robustness}, this paper extends formal verification to more complex RNN architectures. This involves addressing the challenges of the complex structure of the LSTM layers, comprehensively evaluating their behavior, and ensuring robustness compliance through formal verification. This study pushes formal verification's boundaries, embracing design complexities for heightened assurance and reliability.
    \item In this assessment, we conduct a thorough and comprehensive evaluation of three distinct network architectures across diverse audio classification scenarios. 
    \item Finally, we develop insights on evaluating the reachability analysis on those networks and possible future direction.
\end{enumerate}

\paragraph*{Outline.}
The paper is organized as follows:
Section \ref{Sec: relatedwork} mentions the works already done in the literature and the inspiration works for this paper; Section \ref{Sec: preli} provides the necessary context for the background, 
and defines the verification properties for this work; Section \ref{Sec: ReachabilityLayers} explains the reachability calculations for the LSTM layer; and Section \ref{Sec: experiment} describes the methodology, including dataset, network models, and input perturbations. Section \ref{Sec: evaluation} presents the experimental results, evaluation metrics, and their implications. Finally, Section \ref{Sec: conclusion} summarizes the main findings and suggests future research directions.

\section{Related Work}
\label{Sec: relatedwork}
In recent times, an upsurge of methodologies and tools have arisen to confront the verification complexities inherent in intricate systems like Deep Neural Networks (DNNs), as evident from the literature \cite{hashemi2023neurosymbolic, huang2020survey, liu2021algorithms, tran2020verification}. Correspondingly, tools have emerged to tackle the robustness challenges of Convolutional Neural Networks (CNNs) \cite{anderson2019optimization, katz2019marabou, kouvaros2018formal, ruan2018global, singh2018fast, singh2019abstract}. Earlier undertakings in the verification of Recurrent Neural Networks (RNNs) are showcased through projects like RnnVerify \cite{jacoby2020verifying} and RNSVerify \cite{akintunde2019verification}. RNSVerify employs an unrolling technique to translate RNNs into extensive Feedforward Neural Networks (FFNNs), simplifying verification through Mixed-Integer Linear Program (MILP) approaches \cite{akintunde2019verification}. However, this unrolling method faces scalability constraints, particularly with bounded n-step RNNs, as the verification complexity scales dramatically. Conversely, RnnVerify \cite{jacoby2020verifying} employs invariant inference for RNN verification, bypassing unrolling. Their strategy involves crafting an FFNN with matching dimensions to over-approximate the RNN, followed by verifying RNN properties over this approximation using SMT-based methodologies. Our work gets inspiration from \cite{tran2023verification}, where authors introduce a pioneering approach founded on star reachability for RNN verification, aiming to amplify the dependability and safety of RNNs and show the results based on some vanilla RNN models.

\paragraph{Distinction from the previous works \cite{tran2023verification, pal2023robustness}.}
While both papers share the common goal of validating the robustness of RNNs, the preceding study can be perceived as an initial step in that research trajectory. In contrast, this paper represents a more comprehensive evolution of the concepts initially introduced.
\begin{enumerate}
    \item The work in \cite{tran2023verification} focused on the Vanilla RNN, while this paper delves into models of significantly greater intricacy, such as the LSTM and CNN-LSTM architectures. Vanilla RNNs and LSTMs are both types of recurrent neural networks. However, Vanilla RNNs are simpler in terms of architecture and have fewer parameters, whereas LSTMs are more complex due to their gated units and larger parameters.
    \begin{enumerate}
        \item Vanilla RNNs handle input sequences in a sequential manner, updating a hidden state at each step. In contrast, LSTMs also maintain a hidden state, but their structure is more complex, featuring multiple gates (input, forget, and output gates) that regulate the information flow.
        \item Vanilla RNNs face challenges capturing long-term dependencies within sequences due to the vanishing gradient problem. This problem limits their ability to learn connections between distant time steps. LSTMs, on the other hand, were specifically designed to tackle the vanishing gradient problem and excel at capturing long-term dependencies, rendering them better suited for tasks involving intricate temporal relationships.
        \item Vanilla RNNs possess a constrained memory capacity, often rapidly discarding information from earlier time steps. This limitation can hinder their performance in tasks with extended sequences. In contrast, LSTMs feature an improved memory mechanism that allows them to retain or discard information from prior time steps selectively. This capability equips them to handle longer sequences and capture complex patterns effectively.
    \end{enumerate}
    \item The study conducted in \cite{pal2023robustness} focused on examining time-series regression models in the Prognostics and Health Management domain. Drawing inspiration from this work, our study extends the investigation to encompass the time domain's influence, specifically concerning sequential audio noise and Japanese vowel audio samples.
    This basic experiment provides a foundation for understanding the robustness and reliability of audio classification systems. They offer insights that can be directly applied to real-world scenarios, making them valuable for a broad audience in the field of audio classification.
    \begin{enumerate}
        \item Utilizing real-world datasets, this experiment can yield practical insights into audio classification system performance, benefiting fields such as speech recognition, audio surveillance, and multimedia content management by offering real-world applicability.
        \item This experiment can provide valuable insights into the robustness of audio classifiers when exposed to different noise levels and perturbations, offering crucial implications for applications where audio data is frequently affected by a noise like voice commands in automobiles or audio analysis in noisy settings.
        \item While this work concentrates on two particular datasets, the verification methodologies showcased can be extended to diverse audio classification endeavors, allowing readers engaged in various audio classification challenges to customize and apply the methodologies to their specific contexts.
        \item The metrics used in this paper can be potential for real-world applications to evaluate and enhance the reliability and efficiency of audio classification systems.
    \end{enumerate}
\end{enumerate}

\section{Preliminaries}\label{Sec: preli}
\noindent{This section introduces some basic definitions and descriptions necessary to understand the progression of this paper and the necessary evaluations on audio classification models.}

\subsection{Neural Network Verification Tool and Star Sets}
The Neural Network Verification (NNV) tool constitutes a framework designed to verify the safety and robustness of neural networks \cite{tran2020nnv, lopez2023nnv}. This tool meticulously scrutinizes neural network behavior across diverse input conditions, warranting secure and accurate functionality across all scenarios. NNV employs reachability algorithms, including the exact and over-approximate star set methodologies \cite{tran2020verification, tran2019star}, to compute reachable sets for each neural network layer. These sets encapsulate all feasible network states for a given input, thereby facilitating the verification of specific safety properties.

NNV holds particular significance in safety-critical domains like autonomous vehicles and medical devices, where the trustworthiness and reliability of neural networks are paramount. NNV bolsters public confidence in these applications by ensuring consistent performance across all conditions. In this paper, we extend the capabilities of the NNV tool to implement our work, utilizing the star-based reachability analysis to ascertain the reachable sets of neural networks at their outputs.
\begin{figure}[ht!]
    \centering
     \vspace*{-\baselineskip}
    \includegraphics[width = 0.8\columnwidth]{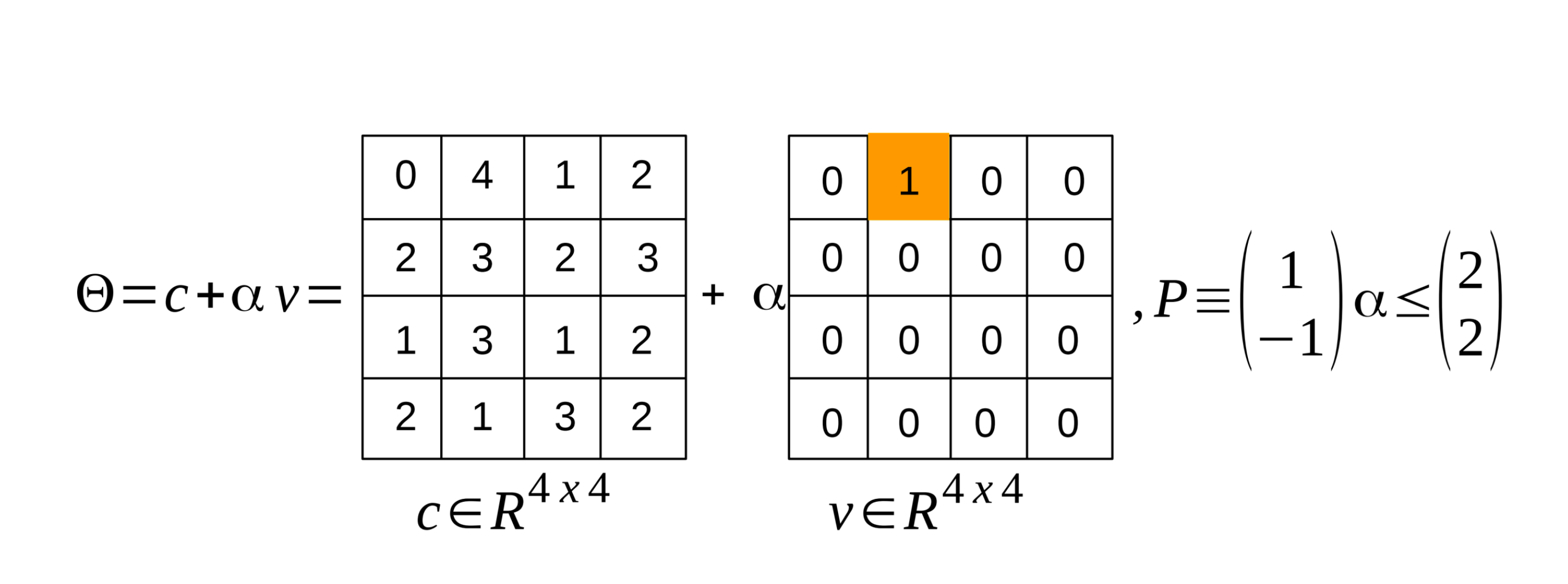}
    \caption{Star for a sequence input data with four Feature Values (rows) with four time-steps (columns)}
    \label{fig:Signalstar}
\end{figure}

\begin{definition}\label{def:star} \emph{\textbf{A generalized star set}} (or simply star) $\Theta$ is a tuple $\langle c, V, P \rangle$ where $c \in \mathbb{R}^n$ is the center, $V = \{v_1, v_2, \cdots, v_m\}$ is a set of m vectors in $\mathbb{R}^n$ called basis vectors, and $P: \mathbb{R}^m \to \{ \top, \bot\}$ is a predicate. The basis vectors are arranged to form the star's $n \times m$ basis matrix. The set of states represented by the star is given as:
\begin{equation}
 \llbracket \Theta \rrbracket = \{x~|~x = c + \Sigma_{i=1}^m(\alpha_iv_i)~\text{and}~P(\alpha_1, \cdots, \alpha_m) = \top \}.
\end{equation}
In this work, we restrict the predicates to be a conjunction of linear constraints, $P(\alpha) \triangleq C\alpha \leq d$ where, for $p$ linear constraints, $C \in \mathbb{R}^{p \times m}$, $\alpha$ is the vector of $m$-variables, i.e., $\alpha = [\alpha_1, \cdots, \alpha_m]^T$, and $d \in \mathbb{R}^{p \times 1}$.
\end{definition}

\vspace*{-\baselineskip}

\subsection{Network Architecture Specifics}
\subsubsection{Long Short Term Memory (LSTM) Layer}
An LSTM layer, a subtype of the Recurrent Neural Network (RNN) layer, excels at capturing long-term dependencies in time series and sequential data \cite{hochreiter1997long}. It comprises two critical elements: the hidden state ($h_t$, also called the output state) and the cell state ($c_t$). At each time step `t,' the hidden state captures the layer's output for that instance, while the cell state accumulates insights from preceding time steps.
\bigskip
\begin{figure}[ht!]
    \centering
     \vspace*{-\baselineskip}
    \includegraphics[width = 0.5\columnwidth]{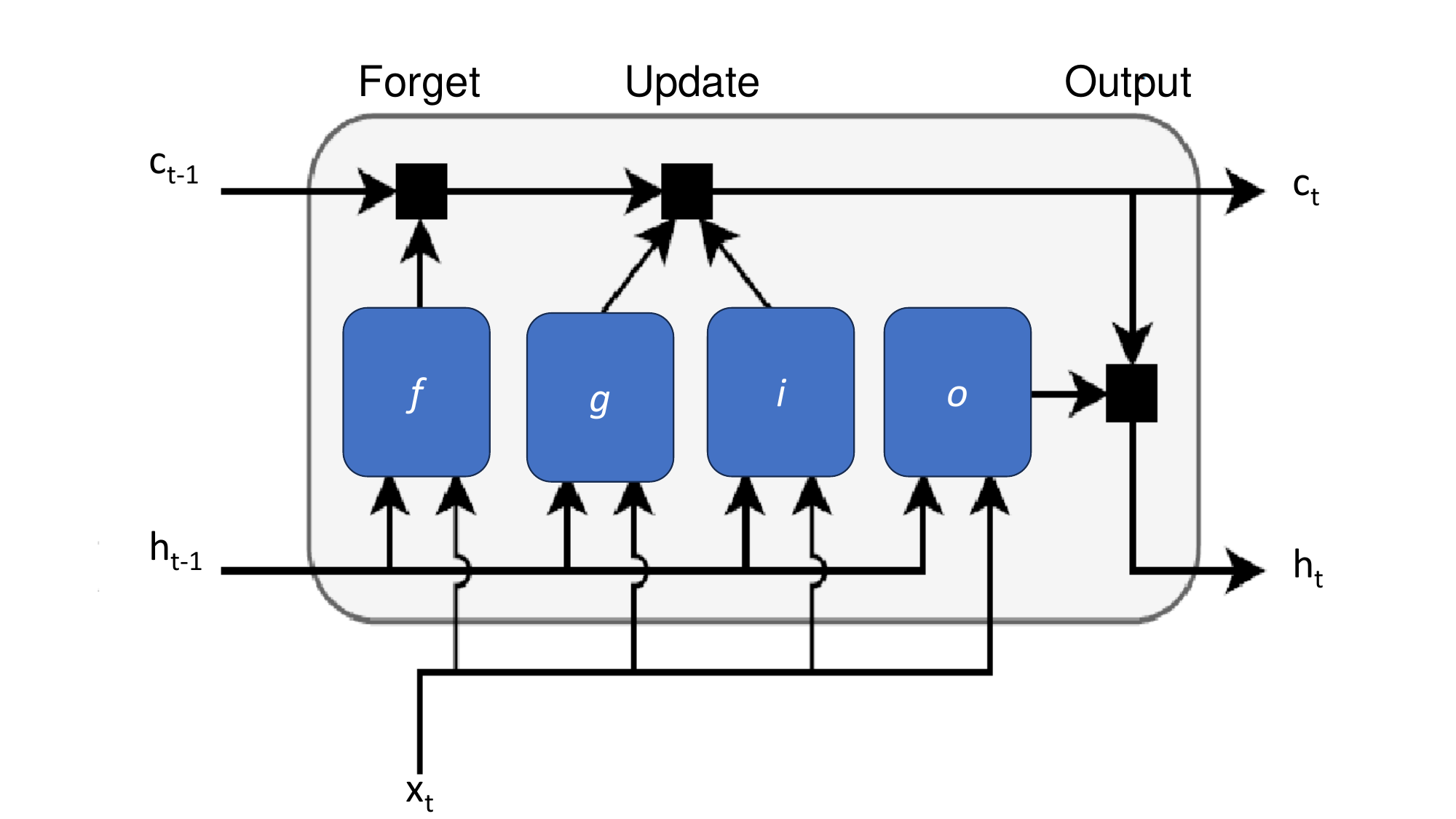}
    \caption{The flow of data at time step t in an LSTM layer}
    \label{fig:lstmgates}
\end{figure}
\vspace*{-\baselineskip}
\begin{equation}\label{eq1}
    \begin{split}
        c_t &= f_t \odot c_{t-1} + i_t \odot g_t \\
        h_t &= o_t \odot \sigma_c(c_t)
    \end{split}
\end{equation}
During each time step, the layer refines the cell state by incorporating or omitting information. This process is steered by distinct gates that control these adjustments, as shown in Fig. \ref{fig:lstmgates}.
\begin{equation}\label{eq2}
    \begin{split}
        i_t = \sigma_g(W_i x_t + R_i h_{t-1} + b_i) \\
        f_t = \sigma_g(W_f x_t + R_f h_{t-1} + b_f) \\
        g_t = \sigma_c(W_g x_t + R_g h_{t-1} + b_g) \\
        o_t = \sigma_g(W_o x_t + R_o h_{t-1} + b_o) 
    \end{split}
\end{equation}
In these equations, \(\odot\) represents the Hadamard product (element-wise multiplication), \(\sigma_c\) denotes the activation function applied element-wise to the cell state \(c_t\) and to the cell state gate \(g_t\); \(\sigma_g\) denotes the activation function applied element-wise to the hidden state gates. Here, $W$, $R$, and $b$ are, respectively, hidden state weights, recurrent weights, and biases for each of the gates.
\subsubsection{Convolutional Neural Network + Long Short Term Memory (CNN+LSTM) Architecture}
When processing sequences, a CNN uses sliding convolutional filters over the input, extracting information from spatial and temporal dimensions. Conversely, an LSTM network progresses through time steps, capturing lasting connections between them. The synergy of CNN and LSTM layers, as seen in CNN+LSTM architectures \cite{zhao2019speech}, harnesses the strengths of both convolutional and LSTM units for insightful data analysis.

The convolutional component forms the foundation for acquiring local feature modules that grasp both local and hierarchical correlations. This fusion enables the identification of intricate data relationships. Additionally, the inclusion of an LSTM layer enhances the network's capacity to capture prolonged dependencies by leveraging information from these localized features.
\begin{figure}[ht!]
    \centering
     \vspace*{-\baselineskip}
    \includegraphics[width = 0.8\columnwidth]{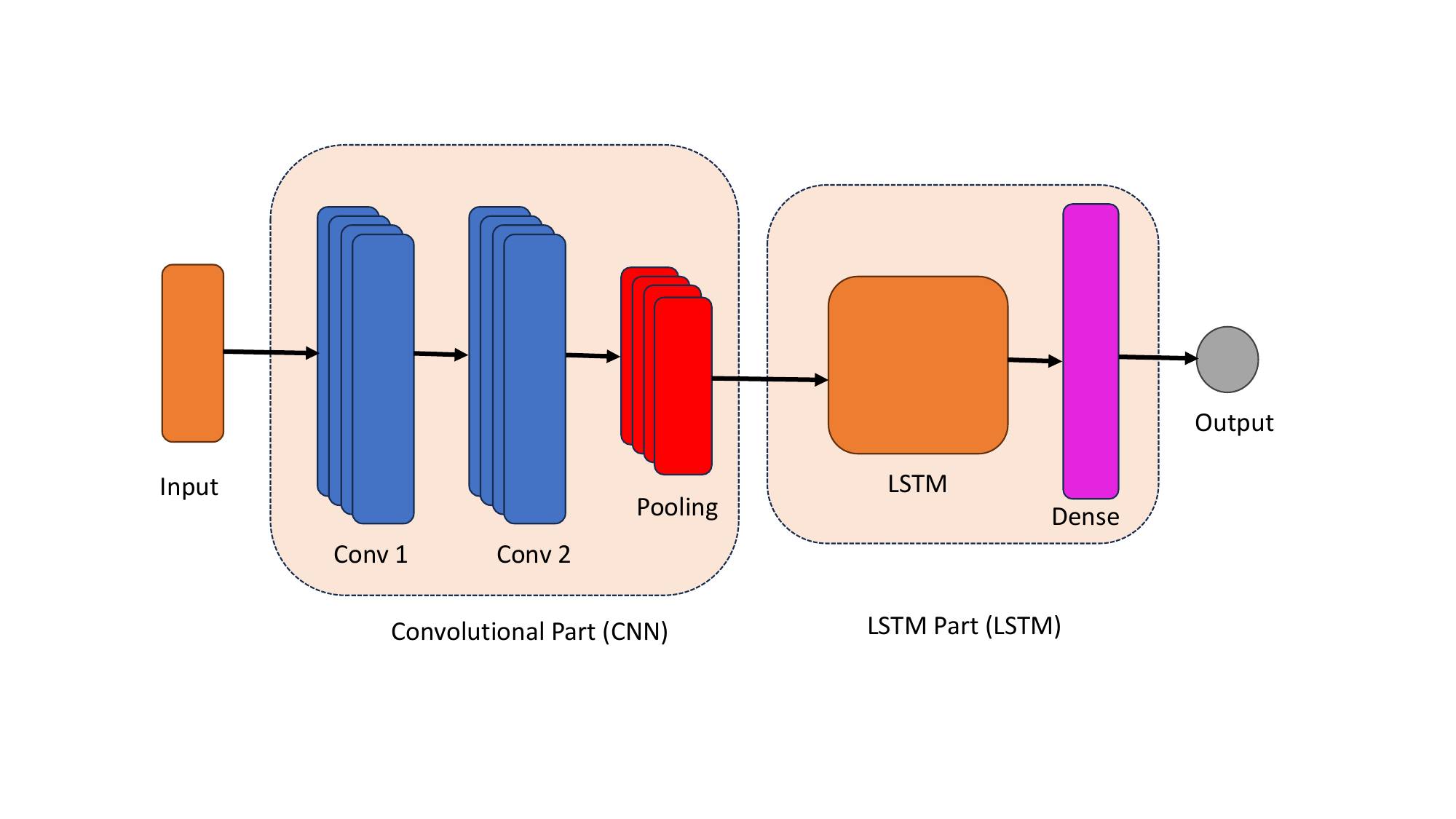}
     \vspace*{-\baselineskip}
    \vspace*{-\baselineskip}
    \vspace*{-\baselineskip}
    \caption{Layers of a demo CNN+LSTM Architecture model}
\end{figure}

\subsection{Reachability Analysis Computation}
This section describes how the reachability of an NN layer and the NN as a whole is computed for this study. 

In this context, we adopt an alternative technique for defining a Star set. This method involves utilizing the input's upper and lower bounds with noise, subsequently aligning them around the original input. We establish a comprehensive array of constraints by incorporating these bounds for each input parameter alongside predicates. These constraints are then presented to the optimizer for a solution, ultimately yielding the initial set of states.

\begin{definition}\label{Def: Layer} A \textbf{layer} $L$ of a NN is a function $h:~u \in \mathbb{R}^{j} \rightarrow v \in {\mathbb{R}}^{p},$ 
defined as
follows 
\begin{equation}
    \label{equ: NNlayer}
    v = h(u)
\end{equation}
where the function $h$ is determined by parameters $\theta$, typically defined as a tuple $\theta = \langle \sigma, W, b \rangle$ for fully-connected
and convolutional layers, where $W$ is the weight matrix, $b$ is the bias vector, and activation function is $\sigma$. For CNN layers, $\theta$ may include parameters like the filter size, padding, or dilation factor.

\end{definition}

\begin{definition} Let $h:~u \in \mathbb{R}^{j} \rightarrow v \in {\mathbb{R}}^{p}$, be an NN layer as described in Eq.~\ref{equ: NNlayer}. The
\textbf{reachable set} ${\mathcal{R}_{h}}$, with input, $\mathcal{I} \in \mathbb{R}^{n}$ is defined as
\begin{equation}
    \label{equ: ReachLayer}
    \mathcal{R}_{h} \triangleq \{v~|~v = h(u),~u \in \mathcal{I}\}
    \end{equation}
\end{definition}

\noindent \textbf{Reachability analysis (or, shortly, reach) of an NN} $f$ on Star input set $\mathcal{I}$ is similar to the reachable set calculations for CNN\cite{tran2020verification} or FFNN\cite{tran2019star}. 
\begin{equation}
\begin{split}
	Reach(f, \mathcal{I}):~&\mathcal{I} \rightarrow \mathcal{R}_{ts} 
\end{split}
\end{equation}
We call $\mathcal{R}_{ts}(I)$ the \emph{output reachable set} of the NN corresponding to the input set $\mathcal{I}$.

For an NN, the output reachable set can be calculated as a step-by-step process of constructing the reachable sets for each network layer. 
 \begin{equation}
    \begin{split}
      \mathcal{R}_{L_{1}} &\triangleq \{v_1~|~v_1 = h_1(x),~x \in \mathcal{I}\}, \\
      \mathcal{R}_{L_{2}} &\triangleq \{v_2~|~v_2 = h_2(v_1),~v_1 \in \mathcal{R}_{L_1}\}, \\
     &\vdots \\
    \mathcal{R}_{ts} = \mathcal{R}_{L_k} &\triangleq \{v_k ~|~ v_k = h_k(v_{k-1}),~v_{k-1} \in \mathcal{R}_{L_{k-1}}\}, \\
    \end{split}
  \end{equation}
  where $h_k$ is the function represented by the $k^{th}$ layer $L_k$. The reachable set $\mathcal{R}_{L_k}$ contains all outputs of the neural network corresponding to all input vectors $x$ in the input set $\mathcal{I}$.

\subsection{Adversarial Perturbation}
An audio classification system may face real-world scenarios involving elements like background noise, interference, or distortions. While potentially perceptible, these factors remain within the scope of challenges that practical systems are designed to address. However, this paper exclusively used l-infinity perturbations, focusing on assessing how audio classification models respond to variations within specific constraints.

Considering an input sequence characterized by $t_s$ time instances and $n_f$ features, various perturbation types ($l_\infty$ norm) \cite{pal2023robustness} arise based on their distribution across the sequence. These adversarial perturbation categories can be delineated as follows:
\begin{enumerate}
    \item \textbf{Single Feature Single-instance (SFSI):} This entails perturbing the value of a specific feature solely at a particular instance ($t$), deviating by a certain percentage from the actual value:
\begin{equation}\label{equ: SFSI}
s^{perturb} = g_{\epsilon, s^{perturb}}(s) = s + \epsilon_t \cdot s_t^{perturb}
\end{equation}
    \item \textbf{Single Feature All-instances (SFAI):} In this scenario, a particular feature across all time instances undergoes perturbation by a certain percentage relative to its original values:
\begin{equation}\label{equ: SFAI}
s^{perturb} = g_{\epsilon, s^{perturb}}(s) = s + \sum_{i=1}^{n}\epsilon_i \cdot s_i^{perturb}
\end{equation}
    \item \textbf{Multifeature Single-instance (MFSI):} All feature values experience perturbation, but solely at a specific instance (t), following the principle outlined in Eq.~\ref{equ: SFSI} for each feature.
    \item \textbf{Multifeature All-instance (MFAI):} Perturbation affects all feature values across all instances, aligning with the approach delineated in Eq.~\ref{equ: SFAI} for every feature.
\end{enumerate}

\subsection{Robustness Verification Properties}
\paragraph{\textbf{Robustness.}}Robustness pertains to the capacity of a system or model to sustain its performance and functionality amid diverse challenging conditions, uncertainties, or perturbations. This highly desirable trait ensures the system's dependability, resilience, and adaptability in the presence of altering or unfavorable circumstances. To formally articulate the concept of robustness for quantifying the desired classification task, the following formulation can be employed:
\begin{equation}
||x' - x||{\infty} < \delta \implies f(x') == f(x)
\end{equation}

\noindent Here, $x$ signifies the original input from the input space $R^{n_f \times t_s}$, $x'$ represents the perturbed input, $f(x')$ and $f(x)$ correspond to the classifiers' outputs for $x'$ and $x$, respectively. $\delta$ stands for the maximum magnitude of the introduced perturbation ($\delta \in \textbf{R} >0$).
By disregarding the softmax and classification layers within the models and focusing on the output of the layer immediately preceding the softmax, the formulation for robustness simplifies as follows:
\begin{equation}\label{Eq: robustclass}
||x' - x||_{\infty} < \delta \implies \text{maxID}(g(x')) == \text{maxID}(g(x))
\end{equation}
In this context, the function $g$ symbolizes the operation performed by the neural network classifier model until the softmax layer, and $\text{maxID}$ denotes the function responsible for identifying the class with the highest value in the output.

\paragraph{\textbf{Verification Properties.}} Verification properties can be broadly classified into two distinct categories: local and global. A local property must be valid for specific predefined inputs, while a global property \cite{wang2022tool} is established across the entire input space $R^{n_f \times t_s}$ of the network model, holding true for all inputs without exceptions. 

\paragraph*{\textbf{Local Robustness.}}\label{Robust} Given a sequence classifier $f$ and an input sequence $S$, the network is called \textbf{locally robust} to any perturbation $\mathcal{A}$ if and only if: reachable bounds of the desired class will be max compared to the bounds of the other classes, even in the presence of any perturbation.

\textbf{Robustness Value (RV)} of a sequence $S$ is a binary variable, which indicates the local robustness of the system. RV is $1$ when the reachable output range of the desired class is greater than the reachable bounds of other classes, making it locally robust; otherwise, RV is 0. 

$RV = 1 \iff {LB_{desired}} \geq {UB_{other}}$
else, RV = 0 

where ${LB_{desired}}$ and ${UB_{other}}$  are the lower reachable bound of the desired class and ${UB_{other}}$ are the upper bounds of all other classes.

\paragraph{\textbf{Percentage Robustness (PR).}\label{def:PR}} We apply the concept of Percentage Robustness (PR), previously utilized in image-based classification or segmentation neural networks \cite{tran2021robustness}, to the context of sequence audio inputs. The PR for a sequence classifier, corresponding to any adversarial perturbation, is defined as:
\begin{equation}
PR = \frac{N_{robust}}{N_{total}}\times 100
\end{equation}
where $N_{robust}$ represents the total number of robust sequences, and $N_{total}$ is the overall count of sequences in the test dataset. Percentage robustness can be used as an indicator of \textbf{global robustness \cite{wang2022tool}} with respect to various types of perturbations.


\section{Reachability of a Long Short Term Memory Layer}\label{Sec: ReachabilityLayers}
To compute the reachability of an LSTM layer in relation to a star input set $S_t$, a series of stepwise reachability computations are necessary to ultimately determine the reachable set of the LSTM layer's output, as depicted in Eq.~\ref{eq1}-\ref{eq2}. Ensuring accurate results relies on verifying the validity of specific conditions, which are crucial for this process to be sound and accurate:
\begin{enumerate}
    \item \textbf{Affine Mapping Validity.} The transformation of a star set through an affine mapping using a given weight and bias must result in another valid star set \cite{tran2019star}.
    \item \textbf{Star Set Summation.} Combining two star sets through Minkowski summation should lead to the formation of yet another valid star set \cite{bak2017simulation}.
    \item \textbf{Activation Function Application.} Upon applying the activation function to a star set, the output should also result in a star set(s). The outcome could manifest as a single star set or a composition of multiple star sets, contingent on factors such as the activation functions employed and the specific reachability technique utilized \cite{tran2019star, tran2020nnv, tran2021robustness, tran2020verification}.
    \item \textbf{Hadamard Product Validity.} The Hadamard Product of two star sets should yield another valid star set.
\end{enumerate}

While the validity of the first three conditions for star sets has been established in prior research, this current study aims to extend that validation to include the fourth condition as well.

\begin{definition}[Hadamard product of two star sets]
Given two star-sets $\Theta_1 = \langle c_1, V_1, P_1\rangle$ and $\Theta_2 = \langle c_2, V_2, P_2\rangle$, the Hadamard product of them $\bar{\Theta} = \Theta_1 \odot \Theta_2 = \{y~|~y=x_1 \odot x_2,~x_1 \in \Theta_1,~x_2 \in \Theta_2 \}$ is another star with the following characteristics.
\begin{equation*}
\bar{\Theta} = \langle \bar{c}, \bar{V}, \bar{P} \rangle, ~ \bar{c} = c_1 \odot c_2, ~ \bar{V} = \begin{bmatrix}
V_1 & 0\\
0 & V_2
\end{bmatrix}, ~ \bar{P} \equiv \begin{bmatrix}
P_1 & 0\\
0 & P_2
\end{bmatrix}
\end{equation*}
\end{definition}

Therefore we can conclude that for a given input set $S_t$ and an LSTM layer, the output is also a star set.

\section{Experimental Setup}\label{Sec: experiment}
\subsection{Hardware Used:} The actual experimental results shown in this paper are conducted in a Windows-10 computer with the 64-bit operating system, Intel(R) Core(TM) i7-8850H processor, and 16 GB RAM.

\subsection{Dataset Description}\label{Dataset}
For evaluation, we consider two different audio datasets for noise classification and Japanese vowel classification.
\paragraph{\textbf{Audio Noise Data}:}
To curate this dataset, we generated a collection of 1000 white noise signals, 1000 brown noise signals, and 1000 pink noise signals using MATLAB. Each signal corresponds to a 0.5-second duration and adheres to a 44.1 kHz sample rate. From this pool of 1000 signals, a training set is fashioned, comprising 800 white noise signals, 800 brown noise signals, and 800 pink noise signals. Given the multidimensionality inherent in audio data, often containing redundant information, a dimensionality reduction strategy is employed. We begin by extracting features and subsequently training the model using only two extracted features. These features are generated from the centroid and slope of the mel spectrum over time.

\paragraph{\textbf{Japanese Vowel \cite{frank2010uci, kudo1999multidimensional}}:}
This dataset is collected from \cite{frank2010uci} from the University of Irvine Machine Learning Repository. Two Japanese vowels were sequentially pronounced by nine male speakers. A 12-degree linear prediction analysis was subjected to each instance of utterances. Each speaker's utterance constitutes a time series ranging from 7 to 29 points in length, with each point featuring 12 coefficients. For 9 classes (i.e., vowels), the dataset has a total of 640 time series. Among these, 270 time series were designated for training purposes, while the remaining 370 were allocated for testing.

\subsection{Network Description}

\paragraph{\textbf{Audio Noise Data}:}
The network architecture used for training the audio noise dataset, partially adopted from \cite{mathworksClassifySound}, is an LSTM network. The network has two input features which correspond to one noise type at the output. Following \ref{Eq: robustclass}, the network for this dataset can be represented as:
\begin{align}
     \label{equ: audionoise}
    \begin{split}
    f:~x \in \mathbb{R}^{2\times l_s} \rightarrow y \in {\mathbb{R}^{3}} \\
        \hat{noiseClass} = maxID(g(x))
    \end{split}
\end{align}

\paragraph{\textbf{Japanese Vowel}:}
Here we have trained two different classifiers for the Japanese Vowel dataset. The LSTM architecture is partially adopted from \cite{mathworksSequenceClassification} and the CNN+LSTM is partially adopted from \cite{mathworksSequenceClassification1d}. Both the networks have twelve input features which correspond to one vowel at the output. Therefore, the networks for this dataset can be represented as:
\begin{align}
     \label{equ: vowel}
    \begin{split}
    f:~x \in \mathbb{R}^{12\times l_s} \rightarrow y \in {\mathbb{R}^{9}} \\
        \hat{vowelClass} = maxID(g(x))
    \end{split}
\end{align}

Here $l_s$ is the audio sequence length and the function $maxID$ provides the class with the maximum value.

\begin{table}[h!]
    \vspace*{-\baselineskip}
    \caption{Performances of different networks used in this paper}
    \label{tab:Table1}
    \centering 
    \begin{tabular}{ccccccc}
    \toprule
    \centering
    $Networks$ & $Accuracy(\%)$\\ 
    \hline 
    $audio\_noise\_lstm$ & 100  \\ 
    $japanese\_vowel\_lstm$ & 93.51 \\
    $japanese\_vowel\_cnnlstm$ & 96.49 \\
    \hline 
    \end{tabular}
\end{table}

\section{Evaluation}\label{Sec: evaluation}

\subsection{Robustness Verification of Audio Noise Classifier}
To conduct robustness verification on the audio noise dataset, we encompass all four categories of perturbations, following \cite{pal2023robustness}. First, we curate 100 sequences each of white, brown, and pink noise as test datasets. Then, we generate adversarial sequences centered around the original ones by applying $l_\infty$ norms. This involves utilizing 5 distinct percentage values for perturbation ($\epsilon$), specifically 50$\%$, 60$\%$, 70$\%$, 80$\%$, and 90$\%$ of the mean ($\mu$) value. These newly created adversarial inputs are subjected to assessment through the exact-star reachability analysis [Sec.~\ref{Sec: ReachabilityLayers}] to determine their robustness. Notably, in the case of Single Feature Single-instance Noise (SFSI) and Single Feature All-instances Noise (SFAI), we opt for random selection of feature 1 for input perturbation.
\begin{table}[h!]
    \caption{Global Robustness: percentage robustness (PR) and total verification runtime (sumRT in seconds) for 100 test audio noise sequences}
    \label{tab:Table2}
    \centering 
    \begin{tabular}{ccccccccc}
    \toprule
    \centering
    $noise$ & $PR_{SFSI}$ & $PR_{SFAI}$ & $PR_{MFSI}$ & $PR_{MFAI}$ & $sumRT_{SFSI}$ & $sumRT_{SFAI}$ & $sumRT_{MFSI}$ & $sumRT_{MFAI}$ \\ 
    \hline 
    50 & 98 & 80.33 & 98 & 80.33 & 0.3071 & 0.2626 & 0.3018 & 0.2625 \\ 
    60 & 96 & 71.67 & 96 & 71.67 & 0.3034 & 0.2571 & 0.3018 & 0.2578 \\
    70 & 94 & 25.67 & 94 & 25.67 & 0.3039 & 0.2637 & 0.3070 & 0.2663 \\
    80 & 85.33 & 12.33 & 85.33 & 12.33 & 0.3073 & 0.2559 & 0.3111 & 0.2556 \\
    90 & 63.33 & 8.33 & 63.33 & 8.33 & 0.3060 & 0.2504 & 0.3093 & 0.2537 \\
    \hline 
    \end{tabular}
    \end{table}
\begin{figure}[h!]
    \centering
    \includegraphics[width = 0.9\columnwidth,]{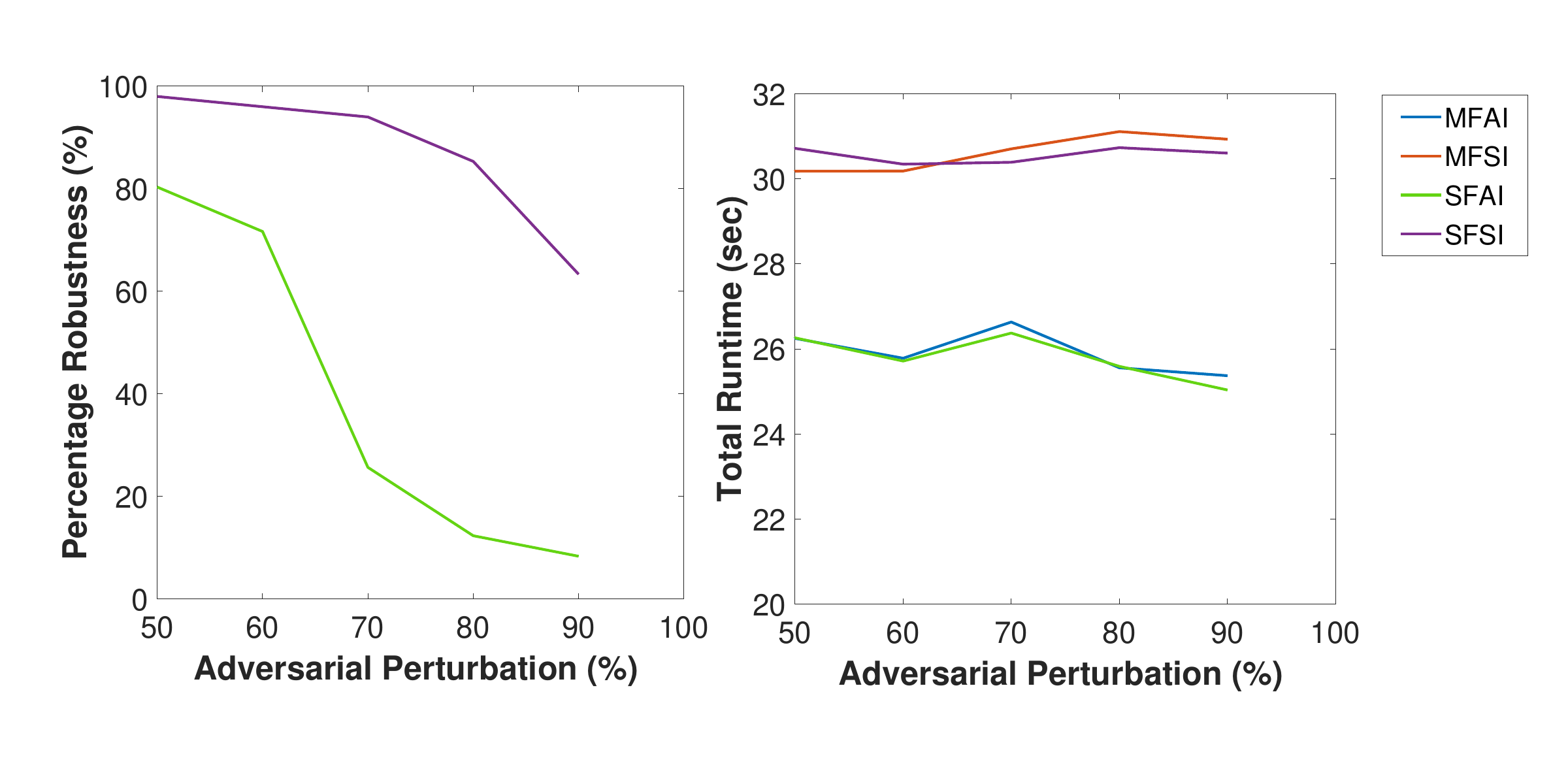}
        \caption{Percentage Robustness and Runtime plots w.r.t increasing perturbations}
    \label{fig:audionoise}
    \vspace*{-\baselineskip}
\end{figure}
\paragraph{\textbf{Observations and Analysis.}} Table~\ref{tab:Table2} and Fig.~\ref{fig:audionoise} present the network's overall performance, i.e., the percentage robustness measures, PR [Sec.~\ref{def:PR}], and total verification runtime (sumRT) in seconds, with respect to each adversarial perturbation. The observations derived from both the table and the figure provide the following insights:
\begin{enumerate}
    \item \textbf{Trend of Percentage Robustness ($PR$).} As the adversary level increases from 50 to 90, we observe a consistent decrease in PR values for all perturbation scenarios (SFSI, SFAI, MFSI, MFAI), which aligns with the concept of the robustness verification property. This decrease in PR signifies a reduction in the system's ability to maintain its classification accuracy in the presence of higher adversary levels.
    \item \textbf{Comparative Analysis of Perturbation Scenarios.} Within each noise level, comparing PR values across different perturbation scenarios (SFSI, SFAI, MFSI, MFAI), it's evident that PR values for SFSI and MFSI are generally higher than those for SFAI and MFAI. This finding indicates that perturbing features at a single instance or all features at a single instance generally leads to better robustness against varying noise levels.
    \item \textbf{Similar PR Values for Different Perturbation Scenarios.} Another notable observation is the similarity in robustness matrices between SFSI and MFSI scenarios, accompanied by closely comparable computation times for their respective verification processes. This parallelism is also evident for SFAI and MFAI perturbations as well. This pattern could be ascribed to the dataset's limited feature set of only two dimensions, where the foremost feature likely holds paramount importance in influencing the class determination in the presence of noise. Consequently, when single-instance perturbations target the first feature, perturbing both features results in an effect akin to perturbing the first feature alone. This interpretation is applicable to both MFAI and SFAI scenarios as well.
\end{enumerate}

\subsection{Robustness Verification of Japanese Vowel Classifiers}
To verify the robustness of both the LSTM and the CNN+LSTM models in the context of the Japanese vowel classifier, we extend the evaluation to encompass all four perturbation categories, mirroring the approach undertaken for the audio noise classifier. During this procedure, we focus on the complete set of correctly classified test sequences. Subsequently, we create adversarial inputs centered around the original sequences by applying $l_\infty$ norms to evaluate their robustness. This perturbation process involves applying 5 distinct percentage values ($\epsilon$) for perturbation: specifically, 50$\%$, 60$\%$, 70$\%$, 80$\%$, and 90$\%$ of the mean ($\mu$) value. The resulting set of adversarial inputs is then assessed using the exact-star reachability analysis for both the classifiers to ascertain their robustness. Like the earlier scenario, for SFSI and SFAI, feature 1 is chosen for perturbation. 


\begin{table}[h!]
    \caption{Global Robustness:  percentage robustness (PR) and total verification runtime (sumRT in seconds) for all test Japanese Vowel audio sequences}
    \label{tab:Table3}
    \centering 
    \begin{tabular}{ccccccccc}
    \toprule
    \centering
    $noise$ & $PR_{SFSI}$ & $PR_{SFAI}$ & $PR_{MFSI}$ & $PR_{MFAI}$ & $sumRT_{SFSI}$ & $sumRT_{SFAI}$ & $sumRT_{MFSI}$ & $sumRT_{MFAI}$ \\ 
    \hline 
    50 & 100 & 68.21 & 100 & 74.86 & 1.1502 & 1.0398 & 1.0392 & 1.0442 \\ 
    60 & 100 & 60.40 & 100 & 50.29 & 0.9989 & 0.9992 & 0.9970 & 0.9966 \\
    70 & 100 & 50.29 & 100 & 27.17 & 0.9981 & 0.9968 & 0.9965 & 0.9952 \\
    80 & 100 & 43.93 & 100 & 13.01 & 0.9920 & 0.9930 & 0.9978 & 0.9882 \\
    90 & 100 & 39.02 & 100 & 8.38 & 1.0044 & 1.0083 & 1.004 & 0.9985 \\
    \hline 
    \end{tabular}
\end{table}
\begin{figure}[h!]
    \centering
    \includegraphics[width = 0.9\columnwidth,]{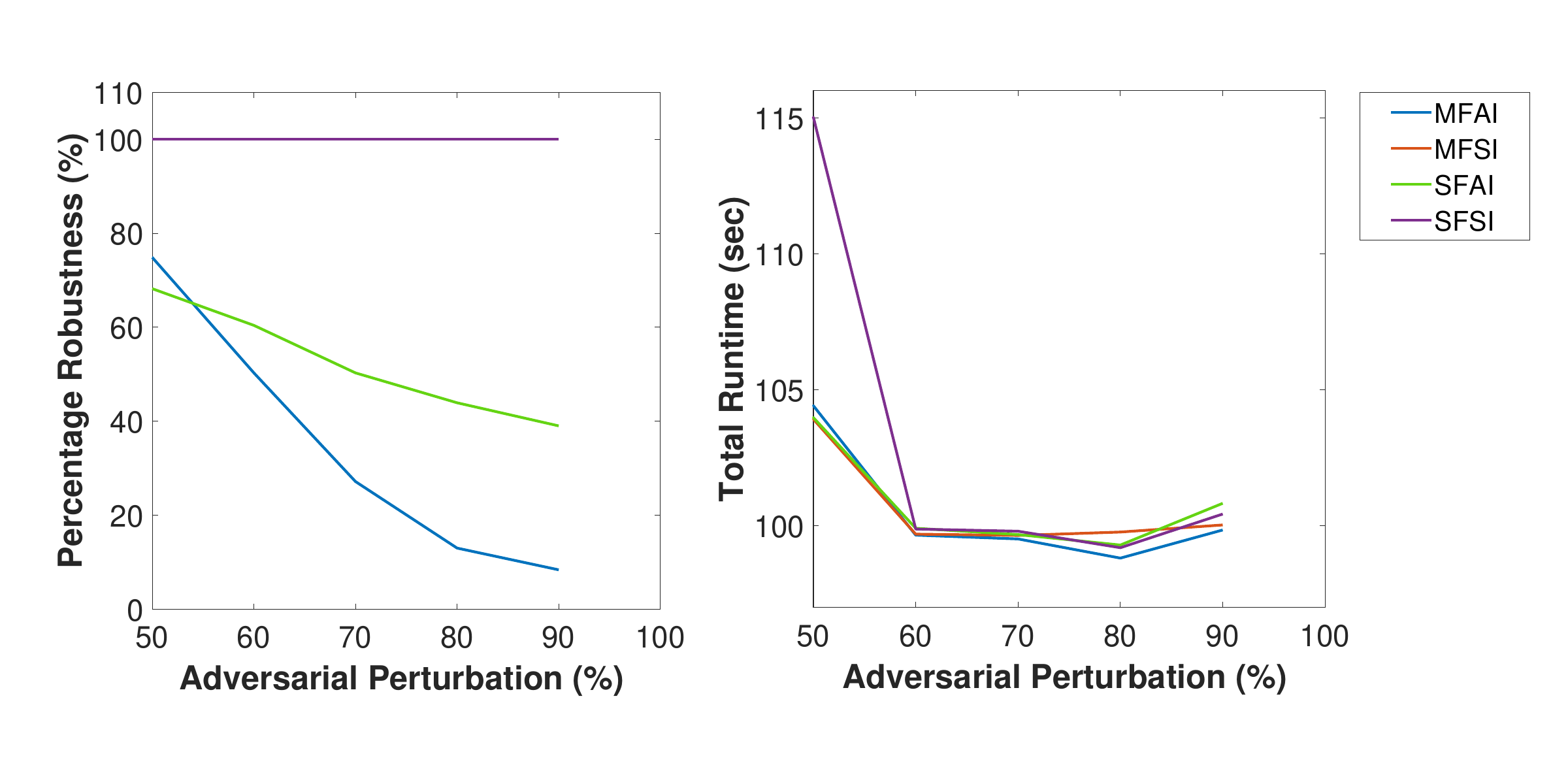}
        \caption{Percentage Robustness and Runtime plots w.r.t increasing perturbations, for LSTM architecture}
    \label{fig:lstm}
    \vspace*{-\baselineskip}
\end{figure}
\paragraph{Observations and Analysis: LSTM Model} 

\begin{figure}[ht!]
    \centering
    \includegraphics[width = 0.9\columnwidth,]{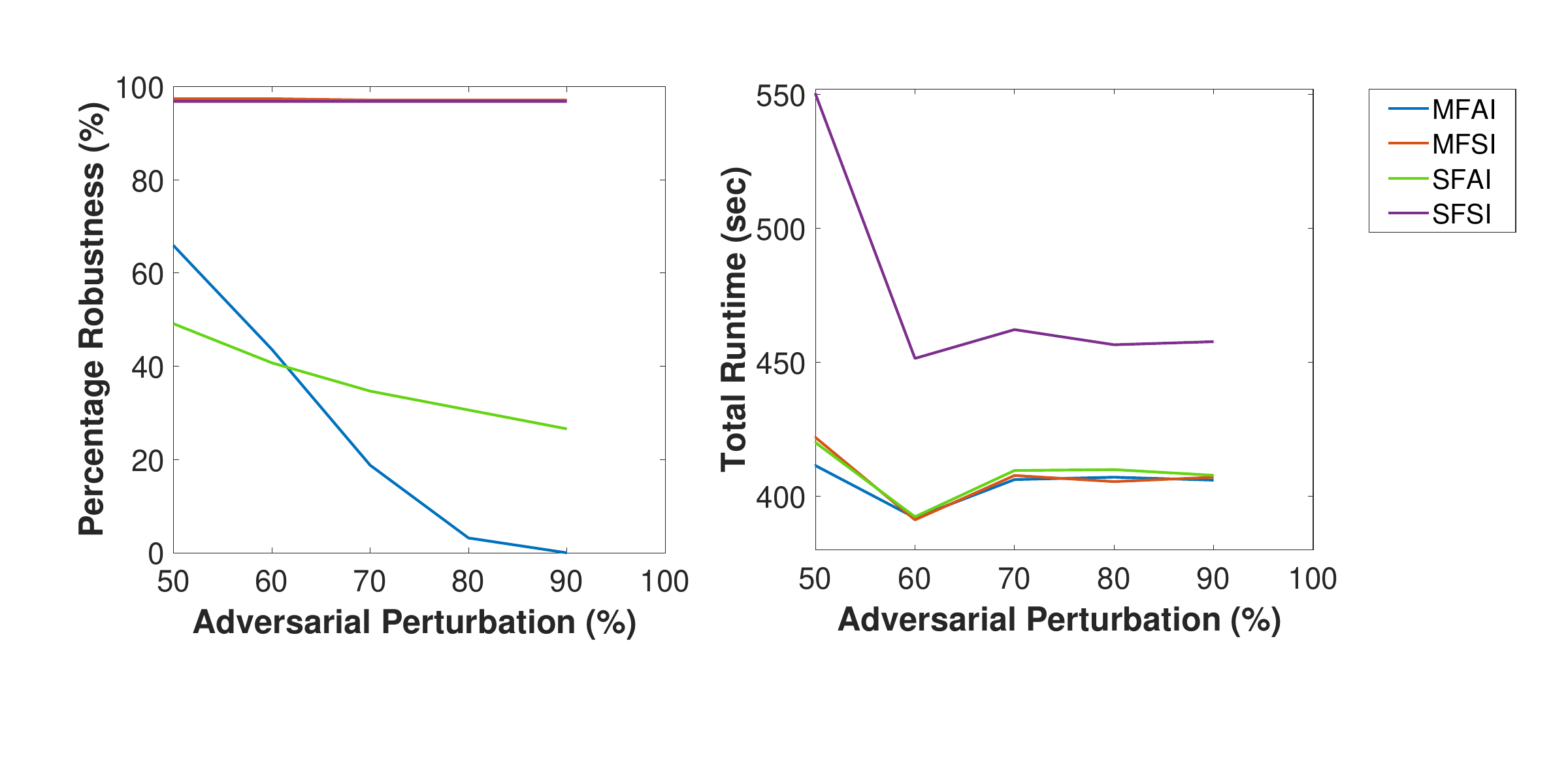}
    \vspace*{-\baselineskip}
        \caption{Percentage Robustness and Runtime plots w.r.t increasing perturbations, for LSTM architecture}
    \label{fig:cnnlstm}
\end{figure}

Table~\ref{tab:Table3} and Fig.~\ref{fig:lstm} present the LSTM network's overall performance, i.e., the percentage robustness measures, PR [Sec.~\ref{def:PR}], and total verification runtime (sumRT), with respect to each adversarial perturbation. The notable findings are outlined as follows
\begin{enumerate}
    \item \textbf{Trend of Percentage Robustness ($PR$).}  Similar to the audio noise classifier, the trends in $PR$ values here also suggest that as noise levels increase, the percentage robustness tends to decrease across all scenarios. This aligns with the intuitive expectation that higher adversary levels lead to increased challenges in maintaining robustness.
    
    The $PR_{SFSI}$ and $PR_{MFSI}$ values remain consistently at 100$\%$ across all noise levels, indicating that perturbing either a single feature or all features at a specific instance does not significantly affect the robustness of the audio sequences. On the other hand, $PR_{SFAI}$ and $PR_{MFAI}$ show distinct trends. As adversary levels increase, $PR_{SFAI}$ gradually decreases, suggesting that perturbing all instances but only a single feature starts impacting the robustness. Similarly, $PR_{MFAI}$ also experiences a decline with increasing noise levels, reflecting that perturbing all instances and features has an impact on the sequences' robustness.

    \item \textbf{Comparative Analysis of Perturbation Scenarios.} The comparison between single-instance perturbation scenarios ($SFSI$ and $SFAI$) and multifeature perturbation scenarios ($MFSI$ and $MFAI$) reveals a pattern. The former scenarios (single-instance) generally maintain higher robustness compared to the latter (multifeature) scenarios. This suggests that perturbing all features has a larger impact on robustness than perturbing just a single feature.
    
    The interrelation between $PR_{SFSI}$ and $PR_{MFSI}$ is also notable. Both scenarios exhibit identical trends, regardless of the noise level. Similarly, $PR_{SFAI}$ and $PR_{MFAI}$ also demonstrate similar behaviors, with both scenarios showing a decline in robustness as noise increases.

\end{enumerate}
\paragraph{Observations and Analysis: CNN+LSTM Model} 
Table~\ref{tab:Table4} and Fig.~\ref{fig:cnnlstm} present the CNN+LSTM network's overall performance.
\begin{table}[h!]
    \caption{Global Robustness:  percentage robustness (PR) and total verification runtime (sumRT in seconds) for all correctly-classified test Japanese Vowel audio sequences}
    \label{tab:Table4}
    \centering 
    \begin{tabular}{ccccccccc}
    \toprule
    \centering
    $noise$ & $PR_{SFSI}$ & $PR_{SFAI}$ & $PR_{MFSI}$ & $PR_{MFAI}$ & $sumRT_{SFSI}$ & $sumRT_{SFAI}$ & $sumRT_{MFSI}$ & $sumRT_{MFAI}$ \\ 
    \hline 
    50 & 96.82 & 49.13 & 97.39 & 65.89 & 5.5019 & 4.2007 & 4.2197 & 4.1148 \\ 
    60 & 96.82 & 40.75 & 97.39 & 43.64 & 4.5148 & 3.9238 & 3.9123 & 3.9200 \\
    70 & 96.82 & 34.68 & 97.10 & 18.78 & 4.6223 & 4.0966 & 4.0778 & 4.0626 \\
    80 & 96.82 & 30.63 & 97.10 & 3.17 & 4.5658 & 4.0998 & 4.0550 & 4.0714 \\
    90 & 96.82 & 26.58 & 97.10 & 0 & 4.5773 & 4.0785 & 4.0715 & 4.0605  \\
    \hline 
    \end{tabular}
\end{table}

Key insights gleaned from both the table and the plot include:
\begin{enumerate}
    \item \textbf{Trend of Percentage Robustness ($PR$).} 
    Across all perturbation levels, the $PR_{SFSI}$ remain consistently at around 96$\%$ and the $PR_{MFSI}$ at around 97$\%$, indicating that the perturbations applied in these scenarios do not significantly affect the robustness of the audio sequences. For SFAI and MFAI perturbations, PR also decreases with rising noise levels, although the decline is more pronounced. PR values for SFSI and MFSI perturbations are significantly higher compared to SFAI and MFAI perturbations at all noise levels, indicating that sequences with perturbations at a single instance are more robust to noise.
    \item \textbf{Trend of Verification Runtimes ($sumRT$).} Verification runtimes tend to rise with elevated noise levels across all perturbation scenarios. However, in the case of the Japanese Vowel dataset, an initial decrease is observed in the runtime trend, followed by an increase at perturbation level 70$\%$ and then again decreases at  80$\%$, followed by another increase at  90$\%$. It's also worth noting that contrary to the expected trend, $sumRT_{SFSI}$ exhibits a higher runtime value in comparison to $sumRT_{SFAI}$ and $sumRT_{MFAI}$.
\end{enumerate}
Overall, the above tables demonstrate how different perturbation scenarios and adversary levels impact the percentage robustness of the audio noise and Japanese Vowel audio classifiers. The trends and interrelations provide insights into the varying effects of perturbations on different scenarios and noise levels, helping to understand the robustness behavior of the neural network models under different conditions.
\section{Conclusion and Future Directions}\label{Sec: conclusion}
This study delves into formal method-based reachability analysis for various LSTM-based neural networks (NNs) using exact and approximate Star methods, specifically in the context of audio sequence classification – a critical aspect for safety-critical applications. The investigation encompasses four distinct adversarial perturbation types, as introduced in the existing literature. The evaluation occurs across two audio sequence datasets: audio noise sequences and Japanese vowel audio sequences. The unified reachability analysis accommodates shifting features within time sequences while scrutinizing the output against the desired audio class. Robustness properties are verified for both datasets. Although real-world datasets are employed, further research is essential to strengthen the connection between practical issues and performance metrics. The evaluation can also be conducted with multiple repetitions to ensure that the reported results are not dependent on specific instances or random fluctuations, thus enhancing the overall validity and reliability of the findings. Exploring real-world scenarios encompassing a wider array of perturbation types and magnitudes will also be fascinating, potentially yielding diverse effects on system behavior. The study paves the way for exploring the impact of perturbations on the output and expanding reachability analysis to three-dimensional sequence data like videos. An intriguing direction for exploration can involve analyzing the peculiar runtime patterns observed in the plots for the Japanese Vowel audio dataset. Potential future applications can also encompass medical video analysis. Notably, this work concentrates on offline data analysis, omitting considerations for real-time stream processing and memory limitations, which offers intriguing prospects for future investigation.

\paragraph{\textbf{Acknowledgements.}}
The material presented in this paper is based upon work supported by the National Science Foundation (NSF) through grant numbers 1910017, 2028001, 2220426, and 2220401, and the Defense Advanced Research Projects Agency (DARPA) under contract number FA8750-18-C-0089 and FA8750-23-C-0518, and the Air Force Office of Scientific Research (AFOSR) under contract number FA9550-22-1-0019 and FA9550-23-1-0135. Any opinions, findings, conclusions, or recommendations expressed in this paper are those of the authors and do not necessarily reflect the views of AFOSR, DARPA, or NSF.

\bibliographystyle{eptcs}
\bibliography{generic}
\end{document}